# $B_s$ Physics

*Thomas Kuhr[1] for the Belle, CDF, and D0 Collaborations*

(1) KIT, Institut für Experimentelle Kernphysik, Wolfgang-Gaede-Str. 1, D76131 Karlsruhe, Germany, Thomas.Kuhr@kit.edu

**Abstract**

While $B^0$ and $B^+$ mesons are well studied, mainly by the B factories, less is known about $B_s$ mesons. Specifically large new physics effects may still be present in the $B_s$ system. This unexplored region is studied by the CDF, D0, and Belle experiments. In this article their recent measurements on the $B_s$ physics sector are presented.

## Introduction

The $B_s$ meson is of particular interest because it is one of the only four kinds of particles that can oscillate into its own anti-particle. Like in the $B^0$ system, this process is described in the standard model (SM) by box diagrams involving the exchange of two W bosons. Deviations from the SM may be observed in oscillation related quantities if new particles contribute to the process. While in the $B^0$ oscillation quarks of first and third generation are connected, leading to large CP violating effects that are precisely measured by the B factories, the $B_s$ system probes transitions between second and third generation quarks where contributions from new physics models are only weakly constrained by experimental results so far. The precise $B_s$ oscillation frequency measurement [1] constrains the absolute mixing amplitude, but not its CP-violating phase. Since the CP violation expected in the $B_s$ mixing is tiny compared to the current experimental resolution, any measurement of a significant CP violation would be evidence for new physics.

In this article, recent measurements of $B_s$ meson decay properties are presented. Not covered are rare decays, like $B_s \rightarrow \mu^+\mu^-$, which are discussed elsewhere in this volume.

## $B_s$ Physics at the Tevatron

The Tevatron proton-antiproton collider is an excellent place to study $B_s$ mesons because of the high b quark production cross section. The b quarks are produced in pairs via strong interaction and the $B_s$ mesons are formed in the fragmentation. The challenges are the high combinatorial background and the huge inelastic cross section which requires highly selective and efficient triggers. Both experiments, CDF and D0, can trigger on dimuons. While D0 has a powerful single muon trigger, the CDF experiment can trigger on displaced tracks to select hadronic B decays.

## $B_s$ Physics at Belle

$B_s$ mesons are produced at the KEKB $e^+e^-$ collider by running at a centre of mass energy of the Y(5S) mass. Since this is slightly more than twice the $B_s^*$ mass the Y(5S) decays in about 20% of the cases to $B_s^{(*)}$ meson-antimeson pairs, where for about 90% of the pairs both states are excited. Besides the Y(5S) branching ratio, the $B_s$ meson yield is further reduced compared to the $B^0$ yield from Y(4S) decays by the lower signal to continuum background ratio. Despite the lower $B_s$ yield than at the Tevatron and insufficient vertex resolution to resolve the $B_s$ oscillations, the Belle experiment can provide complementary measurements, like absolute branching ratios and the observation of decays involving photons that are hard to reconstruct by CDF or D0.

## CP Violation in Semileptonic Decays

CP violation in B mixing can be observed via an asymmetry in "wrong charge" semileptonic decays where the meson oscillates to its anti-particle before it decays. This asymmetry, $a_{SL}$ is, under assumption of CPT invariance, equal to the charge asymmetry of B meson pair decays to like-sign muon pairs, $A_{SL}$. At the Tevatron both kinds of neutral B mesons, $B^0$ and $B_s$, contribute to the overall asymmetry: $a_{SL} = (0.506 \pm 0.043)\, a_{SL}^d + (0.494 \pm 0.043)\, a_{SL}^s$.

To measure the asymmetry, D0 uses a sample of $1.5 \times 10^9$ events with at least one muon candidate, and a dimuon sample of $3.6 \times 10^6$ events [2]. The raw asymmetries are corrected for muons from decay in flight and hadrons mis-identified as muons (called L muons). A further correction for muon detection asymmetries yields the asymmetry for (di)muons from short lived particles (S muons). Taking the fraction of direct semileptonic B decays in these samples from simulation and assuming no asymmetry in the other sources, the final asymmetry for B mesons is derived.

Important ingredients for this analysis are the fractions and asymmetries of L muons and the muon identification asymmetry. Due to the regular reversal of the magnetic fields in the D0 detector the remaining reconstruction efficiency asymmetries are very small. The fractions and asymmetries are determined as a function of $p_T$ (mainly) from data. As a closure test, the asymmetry predicted by these values for the single muon sample, where the S muon contribution is negligible, is compared to the observed asymmetry. As can be seen in Fig. 1 they agree very well.

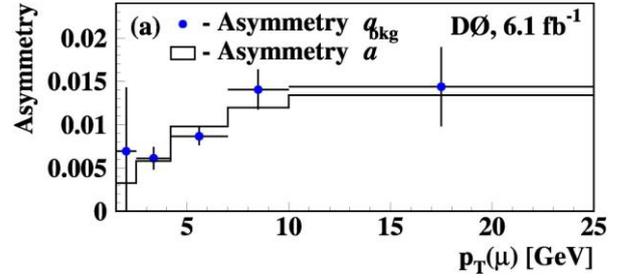

*Fig. 1:* *Comparison of the observed single muon charge asymmetry, a, and the asymmetry predicted by the determined background fractions and asymmetries, $a_{bkg}$.*

An asymmetry of $(+0.94 \pm 1.12 \pm 2.14)\%$ is measured in the single muon sample and of $(-0.736 \pm 0.266 \pm 0.305)\%$ in the dimuon sample where the sensitivity is much higher due to the higher B meson fraction. Since the background asymmetries in both samples are correlated, a further improvement is achieved by combining both measurements to a value of $(-0.957 \pm 0.251 \pm 0.146)\%$. Fig. 2 compares the result to previous measurements.

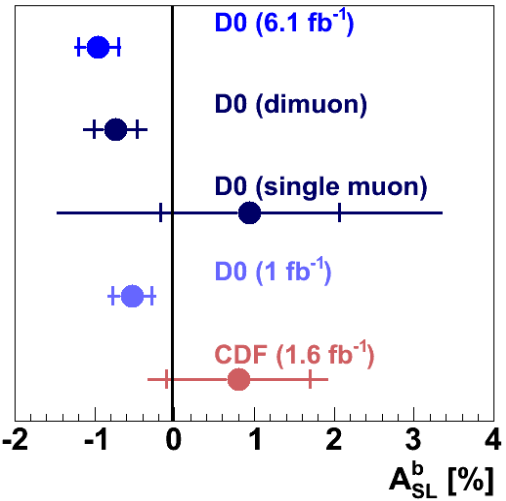

*Fig. 2:* *Semileptonic asymmetry of B mesons for the combined result and the results derived from the dimuon and single muon samples compard to the previous D0 [3] and CDF [4] measurements.*

It was verified that this result is robust against variations of the selection criteria. A visualization of the final result is shown in Fig. 3. The observed dimuon charge asymmetry is consistently below the one that is expected in case of no asymmetry in B decays.

The measured $A_{SL}$ value is 3.2 standard deviations below the value of $(-2.6^{+0.5}_{-0.6}) \times 10^{-4} \approx 0$ expected in the SM [5]. It is therefore considered evidence for an anomalous like-sign dimuon charge asymmetry. Assuming the asymmetry is due to B meson decays, the anomaly may be explained by a new source of CP violation in the $B_s$ or $B^0$ system, or both as illustrated in Fig. 4.

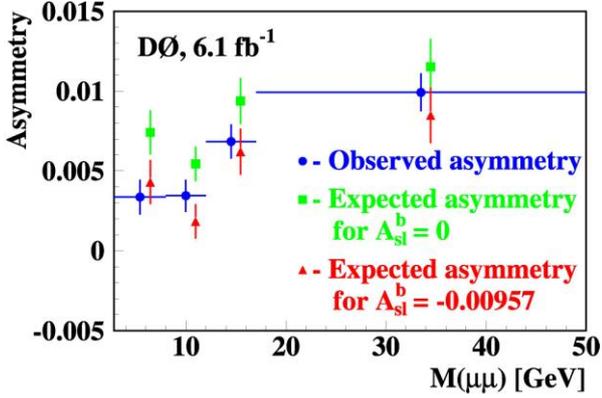

*Fig. 3:* *Observed like-sign dimuon charge asymmetry as a function of dimuon invariant mass compared with the asymmetry expected in case of an asymmetry from B mesons of zero or the measured value.*

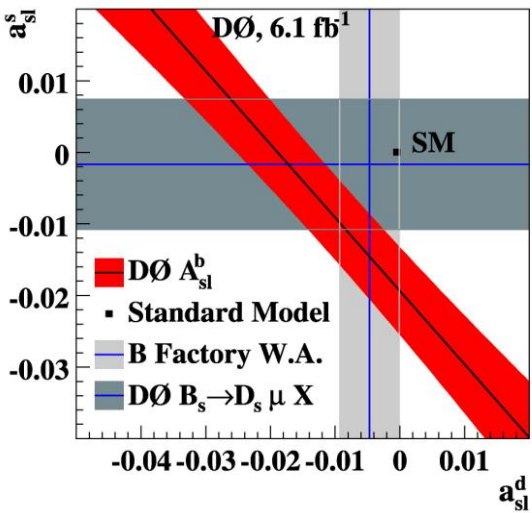

*Fig. 4:* *Constraint of the dimuon charge asymmetry measurement on the semileptonic asymmetries for $B^0$ ($a_{SL}^d$) and $B_s$ mesons ($a_{SL}^s$) compared with direct measurements of these quantities.*

## CP Violation in $B_s \to J/\psi\varphi$

Another way to search for CP violation in the $B_s$ system is a time dependent analysis of decays to a CP eigenstate. The challenges of such a measurement are the resolution of the fast $B_s$ oscillations and the determination of the $B_s$ meson flavour at production (tagging). The decay to $J/\psi\varphi$ is considered the most sensitive one although this final state is a mixture of CP eigenstates and requires an angular analysis. This complication can turn into an advantage because the interference between CP-even and -odd components provides additional sensitivity to CP violation. To exploit this feature the mass eigenstates have to be identified which can only be done if they have different lifetimes.

Both Tevatron experiments have updated their measurements of the CP violating phase in $B_s \to J/\psi\varphi$ decays, $\varphi_s \approx -2\beta_s$. CDF uses ~6500 signal events in a data sample of 5.2 fb$^{-1}$ [6], D0 ~3400 signal events in 6.1 fb$^{-1}$ [7]. The initial flavour of the $B_s$ meson is inferred from the charge of leptons, jets, and vertices produced by the second b hadron from the pair-production (opposite side tag). The taggers which combine this information with a neural network (CDF) or a likelihood ratio (D0) are calibrated on data. In addition CDF exploits the information obtained from particles produced close to the $B_s$ meson in the fragmentation. This same side kaon tagger that was also used for the observation of $B_s$ mixing [1], is for the first time calibrated on data [8]. The amplitude obtained from a mixing fit to a sample of ~13.000 $B_s \to D_s(3)\pi$ decays is well consistent with 1, as can be seen is Fig. 5, indicating a correct calibration. The fitted mixing frequency of (17.79 ± 0.07 (stat)) ps$^{-1}$ agrees well with the published value [1].

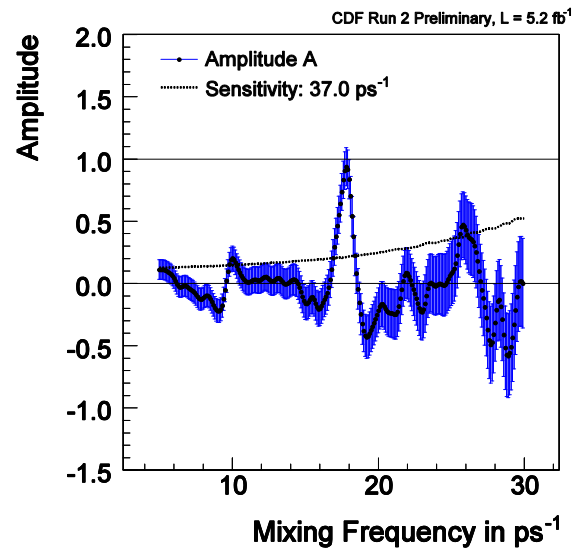

*Fig. 5:* *Fitted amplitude of $B_s$ oscillations as a function of the mixing frequency.*

In previous analyses it was assumed that only J/ψφ decays contribute to the signal yield. Concerns were expressed that there could be a sizable S-wave contamination from $f_0(980) \to K^+K^-$ or non-resonant $K^+K^-$. This would effect the angular distributions and bias the result. Such an effect was observed in $B^0 \to J/\psi K^*(892)$ decays and allowed to resolve a sign ambiguity in the CP violating phase. To address this issue CDF included an S-wave component in the fit. The obtained S-wave contribution is consistent with zero. This is illustrated by the good agreement of $K^+K^-$ mass spectrum with the fit without S-wave component in Fig. 6. D0 looked for an S-wave compnent in their data by measuring the forward-backward asymmetry. The result is consistent with zero as well.

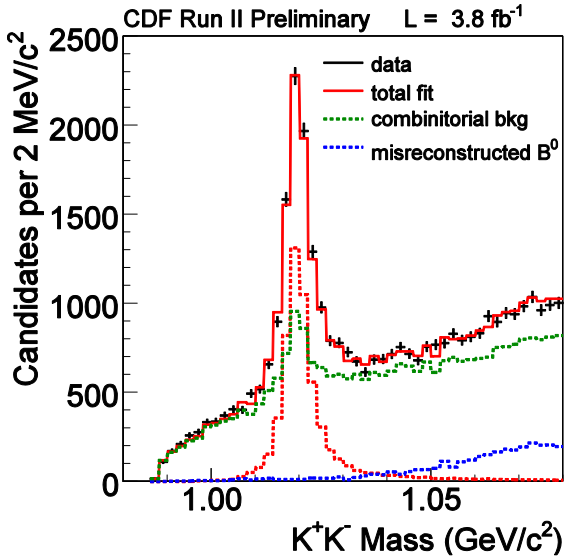

*Fig. 6: Invariant $K^+K^-$ mass distribution with fit assuming no S-wave contribution.*

The final results are shown in Fig. 7 and 8 as 2-dimensional confidence regions in the plane of the CP violating phase and the decay width difference, ΔΓ. The ambiguity observed in the CDF plot is resolved in the D0 plot by a constraint of the strong phases to the ones measured in $B^0 \to J/\psi K^*$ decays. This is motivated by the agreement of the decay amplitudes between both modes.

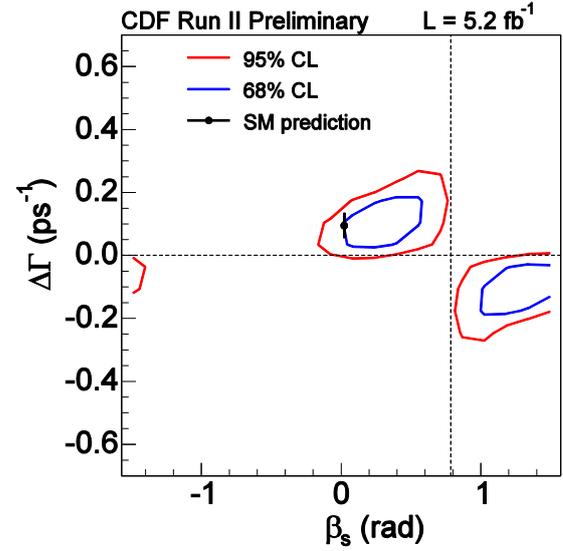

*Fig. 7: Confidence regions for CP violating phase and decay width difference measured by CDF.*

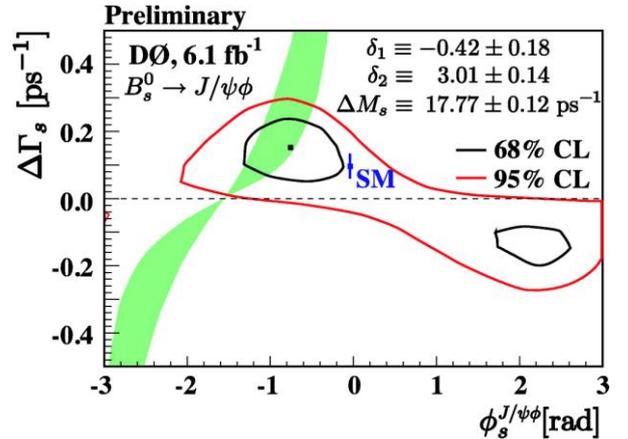

*Fig. 8: Confidence regions for CP violating phase and decay width difference measured by D0. The green band shows the 68% CL region of the $A_{SL}^b$ measurement.*

With the strong phases constraints, D0 is able to quote a central fit value of $\varphi_s = -0.76^{+0.38}_{-0.36} \pm 0.02$. The 68% CL region for $\beta_s$ derived by CDF is [0.02, 0.52] ∪ [1.08, 1.55]. Both results are consistent with each other and with no CP violation and thus agree with the SM.

CDF and D0 also measure the mean lifetime, the decay width difference, the polarization amplitudes, and the strong phase in $B_s \to J/\psi\varphi$ decays. The CDF result is the most precise one today. CDF measures $\Delta\Gamma = (0.075 \pm 0.035 \pm 0.010)$ ps$^{-1}$ and $\tau_s = (1.530 \pm 0.025 \pm 0.012)$ ps.

## $B_s \to \varphi\varphi$

A further interesting mode to search for CP violation is the decay $B_s \to \varphi\varphi$. It is dominated by a $b \to s$ penguin transition. In the SM the mixing induced CP violation is zero and thus searches for new physics are not limited by theoretical uncertainties. Furthermore this mode allows for a comparison between $b \to cc_{bar}s$ and $b \to ss_{bar}s$ transitions where some discrepancy has been observed in the $B^0$ system.

With the displaced track trigger, CDF has collected a sample of ~300 $B_s \to \varphi\varphi$ decays in 2.9 fb$^{-1}$ of data (see Fig. 9). In 2009 CDF had measured the branching ratio BR($B_s \to \varphi\varphi$) = (2.40 ± 0.21 (stat) ± 0.27 (syst) ± 0.82 (BR)) x 10$^{-5}$ where the last error is due to the uncertainty on the branching ratio of the normalization mode [9].

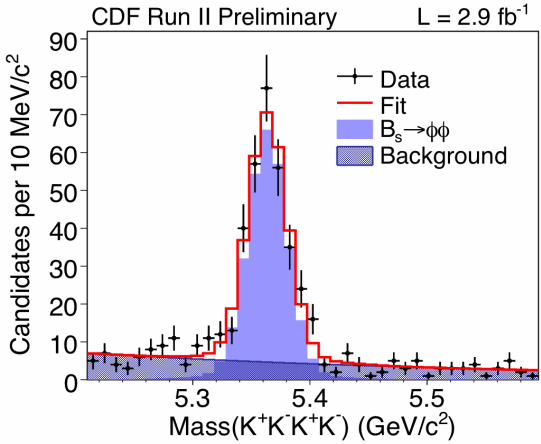

**Fig. 9:** *Reconstructed $B_s \to \varphi\varphi$ mass.*

The next step towards a CP violation measurement is the determination of the polarization. From the V-A nature of the weak interaction and helicity conservation this decay is expected to have dominantly longitudinal polarization. This is confirmed in tree level $B^0$ decays, but violated in $B^0 \to \varphi K^*$ decays. Different explanations for this "polarization puzzle" were proposed, leading to different predictions for the $B_s \to \varphi\varphi$ decay.

In an angular analysis CDF measures a longitudinal polarization of $f_0$ = 0.348 ± 0.041 ± 0.021 in the same sample that was used for the branching ratio measurement [10]. As can be seen in Fig. 10, this result is consistent with an explanation of the polarization puzzle by penguin annihilation (QCDF), while an explanation via final state interaction (pQCD) seems disfavored.

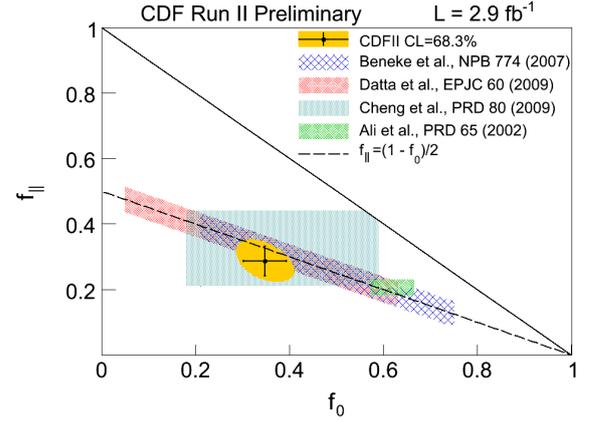

**Fig. 10:** *Polarization measurement by CDF compared with perturbative QCD (Ali et al) and QCD factorization predictions.*

## $B_s \to J/\psi K^*$, and $B_s \to J/\psi K_S$

The SM prediction for CP violation in $B_s \to J/\psi\varphi$ is affected by hadronic uncertainties. The hadronic terms can, with sufficient statistics, be probed by studying the suppressed decay $B_s \to J/\psi K^*$. In a sample of 5.9 fb$^{-1}$ CDF observes a $B_s \to J/\psi K^*$ signal of 151 ± 25 events (see Fig. 11) [11]. With a significance of 8 standard deviations this is the first observation of this decay mode. The branching ratio is measured, relative to the high statistics $B^0 \to J/\psi K^*$ mode, as BR($B_s \to J/\psi K^*$) = (8.3 ± 1.2 (stat) ± 3.3 (syst) ± 1.0 (frag) ± 0.4 (BR)) x 10$^{-5}$. The last two errors are due to the uncertainties on the production ratio of $B_s$ to $B^0$ and the branching ratio of the normalization mode.

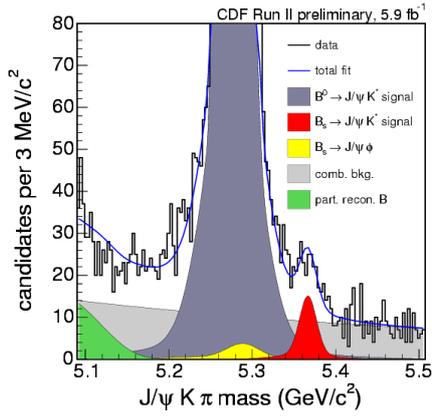

**Fig. 11:** *Invariant $J/\psi K^*$ mass spectrum.*

The suppressed decay $B_s \to J/\psi K_S$ is a CP eigenstate. Therefore it will allow to directly measure the lifetime of the heavy $B_s$ eigenstate if CP and mass eigenstates coincide as expected to very good approximation in the SM. CDF observes a yield of $64 \pm 14$ $B_s \to J/\psi K_S$ events in 5.9 fb$^{-1}$ of data (see Fig. 12). The significance of the signal is 7.2 standard deviations. Thus the decay $B_s \to J/\psi K_S$ is observed for the first time, too. The branching ratio, measured relative to the corresponding $B^0$ decay to the same final state, is BR($B_s \to J/\psi K^0$) = $(3.5 \pm 0.6 \text{ (stat)} \pm 0.4 \text{ (syst)} \pm 0.4 \text{ (frag)} \pm 0.1 \text{ (BR)}) \times 10^{-5}$.

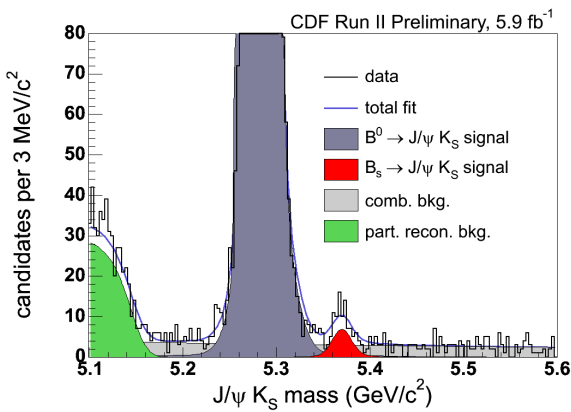

**Fig. 12:** *Invariant $J/\psi K_S$ mass spectrum.*

## $B_s \to J/\psi f_0$

The decay $B_s \to J/\psi f_0$ is considered a promising mode to measure CP violation because it is a CP eigenstate and thus does not require an angular analysis. Belle has searched for this decay in a data sample of 23.6 fb$^{-1}$ of $e^+e^-$ collisions at the $\Upsilon(5S)$ centre of mass energy. From a 2-dimensional fit to the invariant dipion mass spectrum and difference of the reconstructed $B_s$ energy to the beam energy in the centre of mass frame, $\Delta E$, a yield of $6.0 \pm 4.0$ signal events is determined (see Fig. 13). Since no significant signal is observed, a limit of BR($B_s \to J/\psi f_0$) BR($f_0 \to \pi^+\pi^-$) < $1.63 \times 10^{-4}$ at 90% CL is quoted. This agrees with extrapolations from the $B_s \to J/\psi \varphi$ branching ratio [12] and leading order QCD calculations [13].

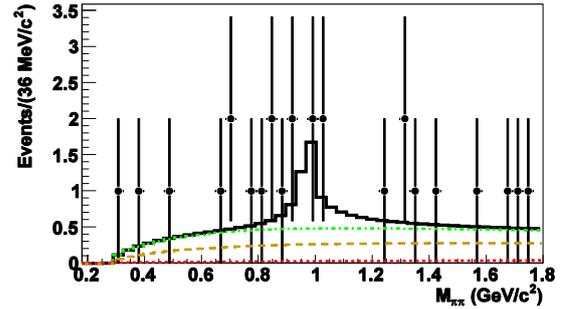

**Fig. 13:** *Invariant $\pi^+\pi^-$ mass spectrum.*

## $B_s \to D_s^{(*)} D_s^{(*)}$

The sensitivity to the CP violating phase in the $B_s \to J/\psi \varphi$ analysis strongly depends on the value of the decay width difference, $\Delta\Gamma$. The difference is caused by decay modes that are accessible to one of the $B_s$ eigenstates, but not to the other one. The main contribution to the width difference between CP-even and -odd states, $\Delta\Gamma_{CP}$, comes from the decay $B_s \to D_s^{(*)} D_s^{(*)}$. It is Cabibbo favored and predominantly CP-even.

In a data sample of 23.6 fb$^{-1}$, Belle has first observed the decay $B_s \to D_s D_s^*$ (6.6$\sigma$, see Fig. 14) and seen first evidence for $B_s \to D_s^* D_s^*$ (3.2$\sigma$) [14]. Branching ratios are measured to be

BR($B_s \rightarrow D_s D_s$) = ($1.0^{+0.4}_{-0.3}{}^{+0.3}_{-0.2}$) %, BR($B_s \rightarrow D_s D_s^*$) = ($2.8^{+0.8}_{-0.7} \pm 0.7$) %, and BR($B_s \rightarrow D_s^* D_s^*$) = ($3.1^{+1.2}_{-1.0} \pm 0.8$) %. Under the assumption that the decay $B_s \rightarrow D_s^{(*)} D_s^{(*)}$ is fully CP-even and no other modes contribute to $\Delta\Gamma_{CP}$ Belle derives $\Delta\Gamma_{CP}/\Gamma$ = ($14.7^{+3.6}_{-3.0}{}^{+4.4}_{-4.2}$) %. This agrees well with the SM prediction [5] and is about two times higher, but still consistent with the D0 result [15].

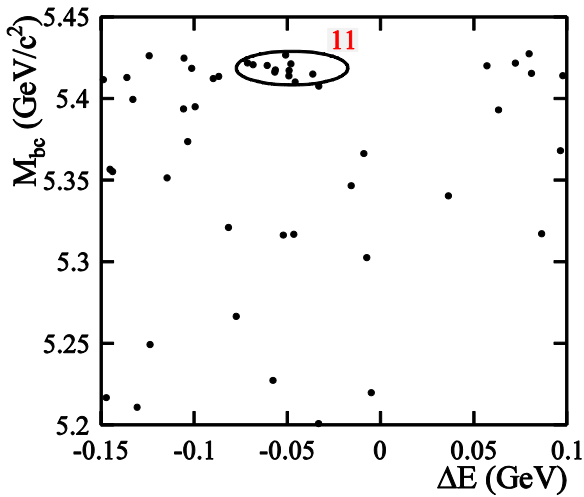

**Fig. 14:** $B_s \rightarrow D_s D_s^*$ *signal peak in the plane of $\Delta E$ and reconstructed B meson mass.*

## Outlook and Summary

Large new physics effects may still be present in the $B_s$ system. This might be an explanation for the anomalous like-sign dimuon charge asymmetry observed by D0. On the other hand the 1.8σ deviation observed in the previous CDF analysis of $B_s \rightarrow J/\psi\varphi$ decays has vanished with more data. The updated result agrees with the SM at a level of 0.8σ. The consistency with the SM has increased in the updated $B_s \rightarrow J/\psi\varphi$ analysis of D0, too, now being at a level slightly above 1σ. The knowledge about the $B_s$ meson system is further increased by measurements of the polarization in $B_s \rightarrow \varphi\varphi$ decays, the branching ratios of $B_s \rightarrow J/\psi K^*$, $B_s \rightarrow J/\psi K^0$, $B_s \rightarrow D_s^{(*)} D_s^{(*)}$, and a limit on BR($B_s \rightarrow J/\psi f_0$).

CDF, D0, and Belle will present updated and new measurements with nearly twice, or even five times in case of Belle, more data at the end of next year. Promising first $B_s$ signals have been shown with early LHCb data. With 1 fb$^{-1}$ LHCb expects to measure the CP violating phase in $B_s \rightarrow J/\psi\varphi$ with about ten times higher precision than the Tevtron experiments and to be able to determine whether $B_s$ or $B^0$ mesons, or both, or some other source causes an anomalous like-sign dimuon charge asymmetry.

## References

[1] A. Abulencia, et al [CDF Collaboration] (2006): *Observation of Bs-Bsbar Oscillations*, Phys. Rev. Lett. 97, 242003

[2] V.M. Abazov, et al [D0 Collaboration] (2010): *Evidence for an Anomalous Like-Sign Dimuon Charge Asymmetry*, Phys. Rev. D 82, 03200; Phys. Rev. Lett. 105, 081801

[3] V.M. Abazov, et al [D0 Collaboration] (2006): *Measurement of the CP-violation parameter of B0 mixing and decay with ppbar -> mu mu X data*, Phys. Rev. D 74, 092001

[4] CDF Collaboration (2007): *Measurement of CP asymmetry in semileptonic B decays*, CDF note 9015

[5] A. Lenz, U. Nierste (2007): *Theoretical update on $B_s$ anti-$B_s$ mixing*, JHEP 0706, 072

[6] CDF Collaboration (2010): *An Updated Measurement of the CP Violating Phase $\beta_s^{J/\psi\varphi}$ with L=5.2 fb$^{-1}$*, CDF note 10206

[7] D0 Collaboration (2010): *Updated measurement of the CP-violating phase $\varphi_S$ using flavor-tagged decay $B_s \rightarrow J/\psi \varphi$*, D0 note 6098

[8] CDF Collaboration (2010): *Calibration of the Same Side Kaon Tagger using 5.2 fb$^{-1}$ of ppbar Collisions*, CDF note 10108

[9] CDF Collaboration (2010): *Updated Measurement of the $B^0_s \rightarrow \varphi\varphi$ Branching Ratio Using 2.9fb$^{-1}$*, CDF note 10064

[10] CDF Collaboration (2010): *Measurement of the Polarization Amplitudes of the $B^0_s \rightarrow \varphi\varphi$ Decay*, CDF note 10120

[11] CDF Collaboration (2010): *Observation of new suppressed Bs decays and measurement of their branching ratios*, CDF note 10240

[12] S. Stone, L.Zhang (2009): *S-waves and the Measurement of CP Violating Phases in B(s) Decays*, Phys. Rev. D 79, 074024

[13] P. Colangelo, et al (2010): *Bs -> f0(980) form factors and Bs decays into f0(980)*, Phys. Rev. D 81, 074001

[14] S. Esen, et al [Belle Collaboration] (2010): *Observation of $B_s$->$D_s^{(*)+}D_s^{(*)-}$ using $e^+e^-$ collisions and a determination of the $B_s$-$B_s$bar width difference $\Delta\Gamma_s$*, arXiv:1005.5177

[15] V.M. Abazov, et al [D0 Collaboration] (2009): *Evidence for decay B0s -> Ds(*) Ds(*) and a measurement of $\Delta\Gamma^{CP}s/\Gamma s$*, Phys. Rev. Lett. 102, 091801